\documentclass[aps,prl,superscriptaddress,preprintnumbers,twocolumn]{revtex4}

\usepackage{epsf,epsfig,amsmath,amssymb,amsfonts, bbm, mathrsfs}
\usepackage{color}
\usepackage{slashed}

\newsavebox{\ns}
\newsavebox{\dbrane}

\def\be{\begin{equation}}
\def\ee{\end{equation}}
\def\bea{\begin{eqnarray}}
\def\eea{\end{eqnarray}}

\def\Dslash{\,\,{\raise.15ex\hbox{/}\mkern-12mu D}}
\def\Dbarslash{\,\,{\raise.15ex\hbox{/}\mkern-12mu {\bar D}}}
\def\delslash{\,\,{\raise.15ex\hbox{/}\mkern-9mu \partial}}
\def\delbarslash{\,\,{\raise.15ex\hbox{/}\mkern-9mu {\bar\partial}}}
\def\pslash{\,\,{\raise.15ex\hbox{/}\mkern-9mu p}}
\def\calDslash{\,\,{\raise.15ex\hbox{/}\mkern-12mu {\cal D}}}

\newcommand{\vol}{\mbox{vol}}
\newcommand{\dd}{\textrm{d}}
\newcommand{\DD}{\textrm{D}}

 \newcommand{\bk}[1]{\langle#1\rangle}
\newcommand{\f}[2]{\frac{#1}{#2}}
\newcommand{\str}{\textrm{str}}
\newcommand{\ale}[1]{\begin{align}#1\end{align}}
\newcommand{\comm}[1]{\left[#1\right]}
\newcommand{\acomm}[1]{\left\{#1\right\}}
\newcommand{\gcomm}[1]{\left[#1\right\}}
\newcommand{\fn}[1]{\footnote{#1}}
\newcommand{\p}[1]{\left(#1\right)}

\newcommand{\nn}{\nonumber \\}

\def\a{\alpha}

\def\b{\beta}

\allowdisplaybreaks

\begin{document}

\title{A Requiem for $AdS_4 \times \mathbb{C} P^3$ Fermionic self-T-duality}

\preprint{DMUS--MP--16/18}            

\preprint{APCTP Pre2016 - 018}

\vskip 1 cm

 \author{E. \'O Colg\'ain} 
 \affiliation{Department of Mathematics, University of Surrey, Guildford GU2 7XH, UK}
 \affiliation{Asia Pacific Center for Theoretical Physics, Pohang 37673, Korea \\
 Department of Physics, Pohang University of Science \& Technology, Pohang 37673, Korea}
 
 \author{A. Pittelli}
  \affiliation{Department of Mathematics, University of Surrey, Guildford GU2 7XH, UK}

\begin{abstract}
Strong evidence for dual superconformal symmetry in $\mathcal{N} = 6$ superconformal  Chern-Simons theory has fueled expectations that the AdS/CFT dual geometry $AdS_4 \times \mathbb{C} P^3$ is self-dual under T-duality. We revisit the problem to identify commuting bosonic and fermionic isometries in a systematic fashion and show that fermionic T-duality, a symmetry originally proposed by Berkovits \& Maldacena, inevitably leads to a singularity in the dilaton transformation. We show that TsT deformations commute with fermionic T-duality and comment on T-duality in the corresponding sigma model. Our results rule out self-duality based on fermionic T-duality for $AdS_4 \times \mathbb{C} P^3$ or its TsT deformations, but leave the door open for new possibilities.  
\end{abstract}

\maketitle

\setcounter{equation}{0}

\section{Introduction} \label{Introduction}
The AdS/CFT correspondence is best understood as an equivalence between $\mathcal{N} = 4$ super Yang-Mills (sYM) and type IIB string theory on $AdS_5 \times S^5$ \cite{Maldacena:1997re}. New ideas incubated in this unique setting quickly percolate to less familiar forms of the duality, where generality may be tested. A duality between superconformal $\mathcal{N} =6$ Chern-Simons (ABJM) theory and type IIA superstrings on $AdS_4 \times \mathbb{C} P^3$ \cite{Aharony:2008ug} represents the first challenge. 

For AdS$_5$/CFT$_4$, it is well-known that integrability is present in the planar limit of $\mathcal{N}=4$ sYM and also strings on $AdS_5 \times S^5$ \cite{Minahan:2002ve, Bena:2003wd}, allowing one to connect perturbative regimes of both descriptions. Integrability is believed to play a substantial r\^ole in relations between scattering amplitudes and Wilson loops \cite{Alday:2007hr, Brandhuber:2007yx, Alday:2007he, Drummond:2008aq}, as well as the emergence of a hidden superconformal symmetry \cite{Drummond:2006rz}, whose closure with the original superconformal symmetry leads at tree level to Yangian symmetry \cite{Drummond:2009fd, ArkaniHamed:2009dn}, a recognisable integrable structure. Moreover, this so-called ``dual superconformal symmetry" can be traced back to a self-mapping of the geometry $AdS_5 \times S^5$ under a combination of bosonic and fermionic T-dualities \cite{Ricci:2007eq, Berkovits:2008ic, Beisert:2008iq}. 

Subsequent developments for ABJM theory largely parallel $\mathcal{N} = 4$ sYM. Integrability is again a feature of the planar limit \cite{Minahan:2008hf, Bak:2008cp} and string theory on $AdS_4 \times \mathbb{C} P^3$ \cite{Arutyunov:2008if, Stefanski:2008ik, Gomis:2008jt, Sorokin:2010wn}. Moreover, perturbative calculations provide convincing evidence for amplitude/Wilson loop duality \cite{Bianchi:2011rn, Bianchi:2011dg}, dual superconformal \cite{Huang:2010qy, Gang:2010gy, Chen:2011vv, Bianchi:2011fc} and Yangian symmetry \cite{Bargheer:2010hn, Lee:2010du}. The similarities between AdS$_5$/CFT$_4$ and AdS$_4$/CFT$_3$ duality are striking and led to the hope that a self-dual mapping of the geometry $AdS_4 \times \mathbb{C} P^3$ under a combination of bosonic and fermionic T-dualities could also account for the observed perturbative symmetries in ABJM theory. In this letter we show that the Berkovits-Maldacena transformation \cite{Berkovits:2008ic} inevitably leads to a singularity in the dilaton shift, so self-duality based on fermionic T-duality can not work for $AdS_4 \times \mathbb{C} P^3$. 

This result may come as no great surprise, since we have witnessed a number of no-go results of varying rigour \cite{Adam:2009kt,Grassi:2009yj, Adam:2010hh, Bakhmatov:2010fp, Sorokin:2011mj, Dekel:2011qw}. Despite this, none are completely satisfactory. Earlier statements, e. g. \cite{Adam:2009kt, Grassi:2009yj}, fail to take account of the chirality and the results of ref. \cite{Bakhmatov:2010fp}, whose approach we follow, fail to be comprehensive in exploring all combinations. Finally, ref. \cite{Dekel:2011qw}, which may be viewed as the most robust result, provides only a set of criteria that are sufficient for self-duality. Over the last year we have become aware of geometries based on exceptional supergroups that violate these criteria, but yet are self-dual \cite{Abbott:2015mla, Abbott:2015ava}. The goal of this letter is to revisit $AdS_4 \times \mathbb{C} P^3$ self-duality in the light of this recent result. 

As stated, our approach mirrors ref. \cite{Bakhmatov:2010fp}, but we strive to be more transparent. In contrast, we make no assumption about the nature of internal bosonic T-dualities required for self-duality and will instead drill down on fermionic T-duality in the supergravity limit \cite{Berkovits:2008ic}, where we know six commuting fermionic T-dualities are required. We will show that regardless how the requisite T-dualities are selected, the dilaton shift resulting from the combined fermionic T-dualities is inevitably singular. One may add undetermined constants to remove the singularity, but there will be a mismatch in the scaling of the $AdS_4$ radial direction required for self-duality. We further demonstrate that deformations based on TsT \cite{Lunin:2005jy} do not change this result. This eliminates an immediate and obvious loophole. 

 Finally, although our findings are confined to the supergravity description, we comment on self-duality based on T-duality from the perspective of the sigma model representation of $AdS_4 \times \mathbb{C} P^3$.  We first  write down  the $AdS_4 \times \mathbb{C} P^3$ coset action in the fashion suitable for T-duality: this is common to all sigma models with OSp symmetry; indeed, it is analogous to the one used  in  \cite{Abbott:2015mla} to show self-duality of $AdS_2\times S^2\times S^2$ and $AdS_3\times S^3\times S^3$. This way, we demonstrate that the singularities affecting the duality transformation have a group theoretical interpretation. Then, we notice that $AdS_4 \times \mathbb{C} P^3$ is a submanifold of $AdS_4 \times G_2\p{\mathbb {R}^8}$, where the Grassmannian $G_2\p{\mathbb {R}^8}$ is the space of all planes  in $\mathbb {R}^8$ passing through the origin. As we shall see, $AdS_4 \times G_2\p{\mathbb {R}^8}$  is a coset  model self dual under Buscher procedure, implying  that the duality transformation links together different $AdS_4 \times \mathbb{ C} P^3$ slices of  $AdS_4 \times G_2\p{\mathbb {R}^8}$. However, such a transformation has a non-trivial super-Jacobian and it is unclear whether it can be identified as a quantum symmetry of the sub-model $AdS_4 \times \mathbb{C} P^3$.

\section{Methodology} \label{Methods}
We follow parallels with geometries $AdS_p \times S^p$, $p = 2, 3, 5$ \cite{Berkovits:2008ic, Beisert:2008iq,Dekel:2011qw, OColgain:2012ca}, and $AdS_q \times S^q \times S^q$, $q=2, 3$ \cite{Abbott:2015mla, Abbott:2015ava}, which are known to be self-dual under a combination of bosonic and fermionic T-dualities. We can map $AdS_4$ back into itself by choosing a Poincar\'e metric, 
\be
\label{Poincare_metric}
\dd s^2 = \frac{\eta_{\mu \nu} \dd x^{\mu} \dd x^{\nu}+ \dd r^2}{r^2}, \quad \mu = 0, 1, 2, 
\ee
performing Abelian bosonic T-dualities along the flat coordinates $x^{\mu}$, before finally inverting the radial coordinate $r \rightarrow r^{-1}$. Note that in performing an odd number of bosonic T-dualities we have switched the chirality of the theory from IIA to IIB, so self-duality requires \textit{at least} one additional T-duality along $\mathbb{C} P^3$. We return to this point in due course. 

One direct consequence of employing the Buscher procedure \cite{Buscher:1987sk} is the dilaton $\Phi$ wanders,  
\be
\label{dilaton_shift}
\delta \Phi = 3 \log r,
\ee
with the raison d'\^etre of fermionic T-duality being to reverse this shift, while at the same time restoring the Ramond-Ramond sector to it original incarnation. In the supergravity description, this feat fermionic T-duality accomplishes by inducing a compensating dilaton shift \cite{Berkovits:2008ic}
\be
\delta \Phi = \frac{1}{2} \log \det C, 
\ee
where $C_{ij}$ is a matrix satisfying
\bea
\label{C_diff}
\partial^{\mu} C_{ij} = \bar{\epsilon}_i \Gamma^{\mu} \Gamma_{11} \epsilon_j, 
\eea
with $\bar{\epsilon}_i \equiv \epsilon_i^{T} \Gamma^0$ and $\epsilon_i$ denoting Killing spinors. To ensure that the fermionic isometries commute, the Killing spinors are subject to the constraint \cite{Berkovits:2008ic}
\be
\label{C_constraint}
\bar{\epsilon}_i \Gamma^{\mu} \epsilon_j = 0 \quad \forall~~ i, j. 
\ee
It is worth noting that in the Majorana-Weyl representation for the gamma matrices, where Killing spinors should be real, the bilinear $\bar{\epsilon}_i\Gamma^{\mu} \epsilon_j$ reduces to $\epsilon_i^{\dagger} \epsilon_j =0$ and a solution only exists when the real Killing spinors are complexified. The usual rule of thumb is that if the original geometry preserves $2n$ Killing spinors, we can construct $n$ fermionic isometries through complexification. We are not aware of any exceptions. In this letter we will show that this transformation, applied to $AdS_4 \times \mathbb{C} P^3$, results in a singularity regardless of how the fermionic directions are chosen. 

It is now well documented how T-duality, both Abelian and non-Abelian, affects supersymmetry at the level of supergravity. It can be shown, e. g. \cite{Kelekci:2014ima}, that supersymmetry breaking is encoded in the Kosmann spinorial Lie derivative \cite{Kosmann}
\be
\label{Kosmann} 
\mathcal{L}_{K} \eta = K^{\mu} \nabla_{\mu} \eta + \frac{1}{8} ( \dd K)_{\mu \nu} \Gamma^{\mu \nu} \eta, 
\ee
where $K$ is the Killing vector, corresponding to the isometry on which we T-dualise, and $\eta$ denotes the Killing spinor. Where $\mathcal{L}_{K} \eta = 0$ is a good projection condition, some supersymmetry survives, otherwise all supersymmetry is broken. Employing this result, we see that T-duality on the $x^{\mu}$ coordinates breaks supersymmetries dependent on $x^{\mu}$, leaving only the so-called Poincar\'e Killing spinors, which depend only on the radial coordinate \footnote{More precisely, compactification of these directions through periodic boundary conditions breaks supersymmetry.}. 

The Poincar\'e Killing spinor is schematically $\eta \sim r^{-\frac{1}{2}} \tilde{\eta}$, where $\tilde{\eta}$ depends on the internal coordinates. From (\ref{C_diff}), we recognise that the components of the matrix $C$ scale as $C_{ij} \sim r^{-1}$, so that $\delta \Phi  \sim - \frac{n}{2} \log r$, where $n$ is the number of fermionic T-dualities one performs. This determines the number of fermionic T-dualities that one should perform, notably six in the case of type II $AdS_4$ geometries. 

One final complication in the self-duality of $AdS_4 \times \mathbb{C} P^3$ is that an odd number of bosonic T-dualities will change the chirality of the theory \footnote{Fermionic T-duality does not affect the chirality.}, so to ensure the final result is a solution to Type IIA supergravity, we require \textit{at least} one internal bosonic T-duality along $\mathbb{C} P^3$. As we have just remarked, any additional T-dualities will further break supersymmetry. Therefore, our approach is simple. $\mathbb{C} P^3$ possesses fifteen Killing isometries, which one can potentially T-dualise. For each isometry, we will enumerate the number of preserved supersymmetries with a view to finding candidate sets of six commuting fermionic isometries that can undo the shift in the dilaton (\ref{dilaton_shift}). Where we find the requisite number of fermionic isometries, we will determine the matrix $C_{ij}$. 

We note that this approach was initially adopted in ref. \cite{Bakhmatov:2010fp}, although only three commuting isometries were examined and a singularity was encountered, but other possibilities were not explored. We will be more systematic. To facilitate easy comparison, we will use the same parametrisation of $\mathbb{C} P^3$, but will opt to solve the Killing spinors directly in 10D. In contrast to ref. \cite{Bakhmatov:2010fp}, we find in each case that the determinant of $C$ only vanishes through a cancellation involving the components $C_{ij}$. 

\section{Internal T-dualities}
\label{sec:internal}
As touched upon earlier, we are required to perform internal bosonic T-dualities in order to preserve the chirality of the theory.  Before proceeding to the analysis of $AdS_4 \times \mathbb{C} P^3$, here we digress with a view to highlighting the utility 
of the Kosmann derivative when identifying the appropriate internal isometries on which we can T-dualise and preserve supersymmetry. We recall that as advocated in ref. \cite{Berkovits:2008ic}, in the case of spheres one should analytically continue the sphere to de Sitter and T-dualise along the flat directions, which have obvious shift symmetries. When these isometries are analytically continued back to the original sphere, one notes that the directions are complex. We will now show that one can recover this unusual choice of isometries directly from the Kosmann derivative in the concrete setting of the type IIB solution $AdS_3 \times S^3 \times T^4$, where we parametrise the three-sphere in nested coordinates, 
\be
\dd s^2 (S^3) = \dd \theta^2 + \sin^2 \theta \dd \phi^2 + \sin^2 \theta \sin^2 \phi \, \dd \psi^2. 
\ee
As usual, the three-sphere possesses six Killing directions:   
\bea
P_1 &=& \partial_{\psi}, \quad P_2 = - \cos \phi \partial_{\theta} + \sin \phi \cot \theta \partial_{\phi}, \nn
P_3 + i P_4 &=& e^{i \psi} \left( \sin \phi \partial_{\theta} + \cos \phi \cot \theta \partial_{\phi} + i \cot \theta \csc \phi \partial_{\psi} \right) , \nn
P_5 + i P_6 &=& e^{i \psi} \left( \partial_{\phi} + i \cot \phi \partial_{\psi} \right). 
\eea

Making use of the Kosmann derivative (\ref{Kosmann}) and the Poincar\'e Killing spinors for the geometry \cite{OColgain:2012ca}, 
\be
\label{KS}
\eta = r^{- \frac{1}{2}} e^{- \frac{\theta}{2} \Gamma^{45} \sigma^1 } e^{ \frac{\phi}{2} \Gamma^{34}} e^{\frac{\psi}{2} \Gamma^{45}} \eta_0, 
\ee
where $\eta_0$ is a constant spinor, a short calculation reveals that 
\bea
\mathcal{L}_{P_5} \eta &=& \frac{1}{2} e^{- \frac{\theta}{2} \Gamma^{45} \sigma_1} e^{\frac{\phi}{2} \Gamma^{34}} e^{\frac{\psi}{2} \Gamma^{45}}  \Gamma^{34} \eta_0, \nn
\mathcal{L}_{P_6} \eta &=& \frac{1}{2} e^{- \frac{\theta}{2} \Gamma^{45} \sigma_1} e^{\frac{\phi}{2} \Gamma^{34}} e^{\frac{\psi}{2} \Gamma^{45}}  \Gamma^{35} \eta_0, 
\eea
where we have omitted $\mathcal{L}_{P_i} \eta, i =1, \dots, 4$ for brevity. 

It is evident that none of the above Kosmann derivatives correspond to a good projection condition on their own, yet we can complexify them, $P^5 + i P^6$, so that $\mathcal{L}_{P^5 + i P^6} \eta  = 0 \Rightarrow ( \Gamma^{45} - i ) \eta_0 = 0$ is a good projection condition. This ensures that half the supersymmetry of the original Killing spinor (\ref{KS}) is preserved. We note that we recover the same projection condition from the combination $P^3 + i P^4$. Taken together, $P^3+i P^4$ and $P^5 + i P^6$ correspond to two complex Killing directions that preserve the same supersymmetries. It can be checked that these are indeed the commuting Killing vectors that result when one analytically continues the three-sphere \cite{OColgain:2012ca}.  

In the next section, we will transplant this analysis to the space $\mathbb{C} P^3$. In contrast to the sphere, one striking difference is that $\mathbb{C} P^3$ is already complex. One can attempt to identify the correct isometry directions by rewriting $\mathbb{C} P^3$ so as to make commuting isometries manifest. This approach was adopted in ref. \cite{Bakhmatov:2010fp}, but we will eschew this approach in favour of a direct analysis of the Kosmann derivative for all the isometries of $\mathbb{C} P^3$. 

\section{$AdS_4 \times \mathbb{C} P^3$}
In order to fix normalisations, we start from the maximally supersymmetric $AdS_4 \times S^7$ solution to 11D supergravity, where we write the $S^7$ as a Hopf-fibration over $\mathbb{C} P^3$. We next perform a $\mathbb{Z}_k$ quotient on the Hopf-fibre, which breaks supersymmetry, before reducing to type IIA supergravity. The resulting solution is \cite{Aharony:2008ug}
\bea
\dd s^2 &=& \frac{R^3}{k} \left( \frac{1}{4} \dd s^2 (AdS_4) + \dd s^2 (\mathbb{C} \textrm{P}^3) \right), \nn
e^{2 \Phi} &=& \frac{R^3}{k^3}, \nn 
F_4 &=& \frac{3}{8} R^3 \vol(AdS_4), \quad  
F_2 = k \, \dd \mathcal{A}, 
\eea
where $R$ and $k$ are constants, $\dd s^2(\mathbb{C} P^3)$ denotes the standard Fubini-Study metric and the one-form potential is 
\be
\mathcal{A} =  \frac{i}{2} \frac{(z_{\alpha} \dd \bar{z}_{\alpha} - \bar{z}_{\alpha} \dd z_{\alpha} )}{(1 + |\mathbf{z}|^2)}. 
\ee 
Following \cite{Pope:1984bd}, we introduce real coordinates
\bea
\label{realz}
z_1 &=& \tan \mu \sin \alpha \sin \frac{\theta}{2} e^{\frac{i}{2} ( \psi - \phi + \chi)}, \nn
z_2 &=& \tan \mu \cos \alpha e^{\frac{i}{2} \chi}, \nn
z_3 &=& \tan \mu \sin \alpha \cos \frac{\theta}{2} e^{\frac{i}{2} ( \psi + \phi + \chi)}, 
\eea
where the ranges of the new coordinates are $0 \leq \mu, \alpha \leq \frac{\pi}{2}, 0 \leq \theta \leq \pi, 0 \leq \phi \leq 2 \pi$ and $0 \leq \psi, \chi \leq 4 \pi$. Through the above rewriting, the metric is recast in the form \begin{widetext}
\bea
\dd s^2(\mathbb{C} \textrm{P}^3) &=& \dd \mu^2 + \sin^2 \mu \biggl[ \dd \alpha^2 + \frac{1}{4} \sin^2 \alpha \biggl( \tau_1^2 + \tau_2^2 +  \cos^2 \alpha \tau_3^2 \biggr)  
+ \frac{1}{4} \cos^2 \mu ( \dd \chi + \sin^2 \alpha \tau_3)^2 \biggr], 
\eea
\end{widetext}
where we have introduced the left-invariant one-forms, $ \dd \tau_{a} = \frac{1}{2} \epsilon_{a}^{~ b c} \tau_{b} \wedge \tau_{c}$. The one-form potential becomes 
\be
\mathcal{A} = \frac{1}{2} \left[ \sin^2 \mu ( \dd \chi + \sin^2 \alpha \tau_3) \right]. 
\ee
The Killing vectors on $\mathbb{C} P^3$ take the form \cite{Hoxha:2000jf}: 
\bea
K_{a} &=& \partial_{z_a} + \bar{z}_{a} \bar{z}_{b} \partial_{\bar{z}_{b}}, \quad \bar{K}_{a} = \partial_{\bar{z}_{a}} + {z}_{a} z_{b} \partial_{z_{b}}, \nn
K_{a b} &=& z_{a} \partial_{z_{b}} - \bar{z}_{b} \partial_{\bar{z}_{a}}, 
\eea
where $a, b = 1, 2, 3$ and repeated indices are summed. We note that various vectors are related through complex conjugation, e. g. $\bar{K}_a = (K_a)^*, K_{ab} = - (K_{ba})^*$, so we only need to later determine the Kosmann derivatives for $K_1, K_2, K_3, K_{11}, K_{22}, K_{33}, K_{12}, K_{23}$ and  $K_{31}$, since the remaining derivatives follow through complex conjugation. 

In order to solve the Killing spinor equations, we adopt the supersymmetry conventions from ref. \cite{Kelekci:2014ima} and introduce a natural orthonormal frame for $\mathbb{C} P^3$: 
\bea
e^{\mu} &=& \sqrt{\frac{R^3}{k} } \frac{1}{2} \frac{\dd x^{\mu}}{r}, \quad e^3 = \sqrt{\frac{R^3}{k} } \frac{1}{2} \frac{\dd r}{r}, \nn
e^4 &=& \sqrt{\frac{R^3}{k} } \dd \mu, \quad 
e^5 = \sqrt{\frac{R^3}{k} } \sin \mu \dd \alpha, \nn
e^6 &=& \sqrt{\frac{R^3}{k} }\frac{1}{2} \sin \mu \sin \alpha \tau_1, \quad 
e^7 = \sqrt{\frac{R^3}{k} } \frac{1}{2} \sin \mu \sin \alpha \tau_2, \nn
e^8 &=& \sqrt{\frac{R^3}{k} }\frac{1}{2} \sin \mu \sin \alpha \cos \alpha \tau_3, \nn
e^9 &=& \sqrt{\frac{R^3}{k} } \frac{1}{2} \sin \mu \cos \mu ( \dd \chi + \sin^2 \alpha \tau_3). 
\eea
Since the bosonic T-dualities along $AdS_4$ will break the superconformal supersymmetries, we need only concern ourselves with their Poincar\'e counterparts. To isolate these, we impose the projection condition 
\be
\label{Poincare_spinor}
\Gamma^{012} \sigma^1 \eta = \eta, 
\ee 
where $\eta$ is a Majorana-Weyl spinor 
\be
\eta = \left( \begin{array}{c} \epsilon_+ \\ \epsilon_-  \end{array} \right),  
\ee
with Pauli matrices acting on $\epsilon_{\pm}$. From the dilatino variation, we identify an additional projection condition on the spinors
\be
\label{32to24} (\Gamma^{49} + \Gamma^{58} + \Gamma^{67} ) i \sigma^2 \eta = \Gamma^3 \eta. 
\ee
Together the two projection conditions (\ref{Poincare_spinor}) and (\ref{32to24}) preserve twelve supersymmetries. Recalling our earlier rule of thumb, this suggests that we should be able to construct six fermionic isometries. 

Solving the remaining differential Killing spinor equation coming from the vanishing of the gravitino variation, we identify the explicit form of the Killing spinor: 
\bea
\label{KSE_IIA}
\eta &=& r^{-\frac{1}{2}} e^{\frac{\mu}{2} (\Gamma^9 i \sigma^2 + \Gamma^{43})} e^{- \frac{\alpha}{2} (\Gamma^{54} + \Gamma^{89})}  e^{\frac{\chi}{4} (\Gamma^{67} + \Gamma^{58} + \Gamma^{49})} \nn &\times& e^{\frac{\psi}{4}(\Gamma^{58} - \Gamma^{67})} e^{-\frac{\theta}{4} (\Gamma^{65}+\Gamma^{78})} e^{\frac{\phi}{4}(\Gamma^{58} - \Gamma^{67})} \eta_0, 
\eea
where $\eta_0$ is a constant spinor satisfying (\ref{Poincare_spinor}) and (\ref{32to24}). It should be noted that the condition (\ref{32to24}) commutes through all the exponentials and therefore can be taken to act directly on the constant spinor $\eta_0$. 

\section{Fermionic isometries}
In this section, we enumerate the possibilities for picking commuting fermionic isometries. In the supergravity description, this boils down to isolating Killing spinors that satisfy the condition (\ref{C_constraint}). To this end, we decompose the constant Killing spinor appearing in (\ref{KSE_IIA}) as 
\be
\label{basis_spinor}
\eta_0 = \xi_{+--} + \xi_{-+-} + \xi_{--+} + \xi_{-++} + \xi_{+-+} + \xi_{++-}, 
\ee
where 
\bea
0 = (\Gamma^{49}\mp i ) \xi_{\pm ab} = (\Gamma^{58} \mp i ) \xi_{a \pm b} = (\Gamma^{67} \mp i ) \xi_{a b \pm}. 
\eea
It is worth noting that each of these basis spinors corresponds to two complex supersymmetries \footnote{This violates the Majorana condition. For our spinors to be Majorana, and necessarily real in our conventions, we should impose the projection conditions $(\Gamma^{49} \mp i \sigma_2) \xi_{\pm a b}= 0$, etc.}, allowing a possibility of 12 fermionic isometry directions before we consider the constraint (\ref{C_constraint}). We also remark that the basis spinors $\xi_{+--}$ and $\xi_{-++}$, etc. are, modulo overall coefficients, related via complex conjugation. 

Slotting these eigenspinors into the Killing spinor, we arrive at a final expression for $\eta$: 
\begin{widetext}
\bea
\label{eta}
\eta &=& r^{-\frac{1}{2} } e^{-\frac{i}{4} \chi} \biggl[ \left( \cos \alpha + \sin \alpha \, \Gamma^{45} \, e^{-\mu \Gamma^{43} } \right) \xi_{+--} 
+ e^{\frac{i}{2} \phi} \left( \cos \frac{\theta}{2} e^{\frac{i}{2} \psi} \left( \cos \alpha e^{\mu \Gamma^{43} } + \sin \alpha \Gamma^{45} \right)  + \sin \frac{\theta}{2} e^{- \frac{i}{2} \psi} \Gamma^{56} e^{\mu \Gamma^{43}}  \right) \xi_{-+-} \nn
&+& e^{- \frac{i}{2} \phi}\left( \cos \frac{\theta}{2} e^{- \frac{i}{2} \psi} e^{\mu \Gamma^{43}} + \sin \frac{\theta}{2} e^{\frac{i}{2} \psi} \, \Gamma^{56} \left( \cos \alpha e^{\mu \Gamma^{43}} - \sin \alpha \Gamma^{45} \right) \right) \xi_{--+} \biggr] \nn
&+& r^{-\frac{1}{2} } e^{\frac{i}{4} \chi} \biggl[ \left( \cos \alpha + \sin \alpha \, \Gamma^{45} \, e^{-\mu \Gamma^{43} } \right) \xi_{-++}  
+ e^{-\frac{i}{2} \phi} \left( \cos \frac{\theta}{2} e^{-\frac{i}{2} \psi} \left( \cos \alpha e^{\mu \Gamma^{43} } + \sin \alpha \Gamma^{45} \right)  + \sin \frac{\theta}{2} e^{\frac{i}{2} \psi} \Gamma^{56} e^{\mu \Gamma^{43}}  \right) \xi_{+-+} \nn 
&+& e^{\frac{i}{2} \phi}\left( \cos \frac{\theta}{2} e^{\frac{i}{2} \psi} e^{\mu \Gamma^{43}} + \sin \frac{\theta}{2} e^{-\frac{i}{2} \psi} \, \Gamma^{56} \left( \cos \alpha e^{\mu \Gamma^{43}} - \sin \alpha \Gamma^{45} \right) \right) \xi_{++-} \biggr]. 
\eea
\end{widetext}. 

Making use of the explicit expression for $\eta$ and the Kosmann derivatives in the appendix, we can identify various isometries of $\mathbb{C} P^3$ that can be T-dualised, while still preserving supersymmetry. The result of this analysis is summarised in Table I. 

\begin{table}[ht]
\begin{center} 
\begin{tabular}{cc}
T-duality Isometry & Preserved supersymmetry \\ 
\hline 
$K_1$ & $\xi_{+--}, \xi_{-+-}, \xi_{--+}, \xi_{++-}$ \\
$K_2$ & $\xi_{+--}, \xi_{-+-}, \xi_{--+}, \xi_{-++}$ \\
$K_3$ & $\xi_{+--}, \xi_{-+-}, \xi_{--+}, \xi_{+-+}$ \\
$K_{11} +K_{22}$ & $\xi_{+--}, \xi_{--+}, \xi_{-++}, \xi_{++-}$\\
$K_{11}-K_{22}$ & $\xi_{-+-}, \xi_{+-+} $\\
$K_{22} +K_{33}$ & $\xi_{+--}, \xi_{-+-}, \xi_{-++}, \xi_{+-+}$\\
$K_{22}-K_{33}$ & $\xi_{--+}, \xi_{++-} $\\
$K_{33}+K_{11}$ & $\xi_{-+-}, \xi_{-+-}, \xi_{+-+}, \xi_{++-}$\\ 
$K_{33} -K_{11}$ & $\xi_{+--}, \xi_{-++} $\\
$K_{12}$ & $\xi_{--+}, \xi_{-+-}, \xi_{-++}, \xi_{+-+}$ \\
$K_{23}$ & $\xi_{--+}, \xi_{+--}, \xi_{++-}, \xi_{+-+}$ \\
$K_{31}$ & $\xi_{+--}, \xi_{-++}, \xi_{-+-}, \xi_{++-}$
\end{tabular} 
\caption{Supersymmetries preserved under $\mathbb{C} P^3$ bosonic T-duality.} 
\end{center} 
\end{table}

We remark that the combinations $K_{aa} \pm K_{bb}$, $b \neq a$ are pure imaginary, so modulo an overall factor, they correspond to real bosonic isometries. As a result, we infer from our table that one can generate supersymmetric $AdS_4$ solutions to type IIB supergravity by T-dualising on these directions. The solutions will preserve eight and sixteen supersymmetries, but it is easy to see they will be singular as one T-dualises on a vanishing cycle. 

As emphasised previously, we are committed to performing six fermionic T-dualities in order to undo the dilaton shift. This entails choosing six complex Killing spinors. We note that the requirement that the fermionic isometries commute (\ref{C_constraint}) has a preference for choosing a spinor, but not its complex conjugate. For example, if one considers a the linear combination $\epsilon = \xi_{+--} + \xi_{-++}$, where one allows for arbitrary complex coefficients in the base spinors, we remark that one can only solve the condition (\ref{C_constraint}) when one of $\xi_{+--}$ or $\xi_{-++}$ vanishes. This constraint is consistent with our expectation that the Killing spinors appearing in the fermionic T-duality must remain complex \cite{Berkovits:2008ic}.  

Based on this observation, natural choices for the six commuting fermionic isometries involve choosing constant Killing spinors, denoted $\eta_0$, which satisfy the projection conditions in Table II. Using the results from Table I, we can also list the bosonic isometries of $\mathbb{C} P^3$ that preserve these Killing spinors under T-duality. We omit projection conditions with the opposite signs, which are related through complex conjugation. 

\begin{table}[ht]
\begin{center} 
\begin{tabular}{cc}
Projection & $\mathbb{C} P^3$ isometries \\ 
\hline 
$(\Gamma^{49} + \Gamma^{58} + \Gamma^{67}) \eta_0 = -i \eta_0$ & $ K_1, K_2, K_3$ \\
$\Gamma^{49} \eta_0 = i \eta_0 $ & $K_{21}, K_2^*, K_{23}$ \\
$\Gamma^{58} \eta_0 = i \eta_0 $ & $K_{32}, K_3^*, K_{31}$ \\
$\Gamma^{67} \eta_0 =  i \eta_0 $ & $K_{13}, K_1^*, K_{12}$
\end{tabular} 
\caption{Neglecting complex conjugates, there are four natural sets of six Killing spinors, which are picked out by the above projection conditions. For each set of six Killing spinors, one can identify 3 commuting Killing vectors that may be T-dualised without breaking the supersymmetries. However, regardless of the choice, the determinant of the induced metric is zero, resulting in a singularity in the dilaton shift under T-duality.} 
\end{center} 
\end{table}

We observe that for each choice of 6 Killing spinors, one can identify 3 commuting bosonic Killing directions, as shown in Table II, yet the determinant of the induced metric is always zero! As a direct consequence, we remark that bosonic T-duality with respect to these directions will also result in a singularity in the dilaton shift. However, before we turn our attention to this singularity, we are confronted with a singularity in the fermionic T-duality. This singularity is deeply troubling, since we require the matrix $C_{ij}$ to be invertible so that the  transformation of the RR sector bispinor, tailored to our IIA conventions \cite{Bakhmatov:2011aa}
\be
\frac{i}{16} e^{\tilde{\Phi}} \tilde{F} = \frac{i}{16} e^{\Phi} F + C_{ij}^{-1} \epsilon_i \otimes \epsilon_j, 
\ee
may be executed \footnote{Indeed, fermionic T-duality may be extended to massive IIA, but only constant matrices have been found to date \cite{OColgain:2012si, Bakhmatov:2015wdr}, which result in trivial transformations.}.  

We will now give explicit expressions for the components of the matrix $C_{ij}$ for the various cases highlighted in Table II. We will show in each case that the determinant of $C$ is zero. Contrary to ref. \cite{Bakhmatov:2010fp}, we find that determinant is not zero irrespective of the values of the components, but in fact depends on the cancellation in the components. 

Before proceeding, we make some comments on notation, before presenting results. Recall that there are 12 complex basis spinors (\ref{basis_spinor}), which are determined up to overall complex constants, $a_i, b_i \in \mathbb{C}$ $i = 1, \dots, 6$. We label the constants corresponding to the spinors, $\xi_{+--}, \xi_{-+-}, \xi_{--+}, \xi_{-++}, \xi_{+-+}, \xi_{++-}$ as $a_1, b_, a_2, b_2, \dots, a_6, b_6$, respectively. The explicit basis spinors can be found in the appendix. 

We start by imposing the first projection condition from Table II, namely $\Gamma^3 \sigma^2 \eta_0 =  \eta_0$, which preserves the spinors $\xi_{+--}, \xi_{-+-}, \xi_{--+}$, so the only non-zero complex coefficients are $a_1, b_1, a_2, b_2, a_3, b_3$. With this choice, it is easy to check that (\ref{C_constraint}) is trivially satisfied, so it imposes no further constraint. Modulo an additive constant, which we will comment on soon, one may integrate (\ref{C_diff}) to identify the non-zero components of the matrix: 
\bea
C_{13} &=& - \sqrt{\frac{R^3}{k}} \frac{ 16 \, a_1 a_2\,  i \bar{z}_1}{r (1+|\mathbf{z}|^2)}, \quad C_{16} =   \sqrt{\frac{R^3}{k}} \frac{16 \, a_1 b_3  i \, \bar{z}_3}{r (1+|\mathbf{z}|^2)}, \nn
C_{24} &=& -\sqrt{\frac{R^3}{k}} \frac{ 16 \, b_1 b_2\,  i \bar{z}_1}{r (1+|\mathbf{z}|^2)} , \quad C_{25} = -  \sqrt{\frac{R^3}{k}} \frac{16 \, b_1 a_3  i \, \bar{z}_3}{r (1+|\mathbf{z}|^2)}, \nn
C_{35} &=&   \sqrt{\frac{R^3}{k}}  \frac{ 16 \, a_2 a_3  i \, \bar{z}_2}{r (1+|\mathbf{z}|^2)}, \quad C_{46} = \sqrt{\frac{R^3}{k}}  \frac{ 16 \, b_2 b_3  i \, \bar{z}_2}{r (1+|\mathbf{z}|^2)}, 
\eea
where we have re-expressed the result in terms of the original Fubini-Study coordinates through (\ref{realz}). 

We observe that the determinant of this matrix, 
\be
\label{detC}
\det C = - (C_{16} C_{24} C_{35} + C_{13} C_{25} C_{46})^2, 
\ee
is zero once evaluated, in line with the findings of ref. \cite{Bakhmatov:2010fp}. However, in contrast to ref. \cite{Bakhmatov:2010fp} we do not find that the matrix $C$ is trivially zero, namely cancellation occurs once the components are evaluated. Although we have picked our spinors to coincide with the basis spinors, it is straightforward to check that the determinant is zero for more general linear combinations. It is also obvious that the result for the projection condition with the opposite sign, i. e. $\Gamma^{3} \sigma^2 \eta_0 = - \eta_0$ follows immediately from complex conjugation and the determinant will be again zero. 

Before moving on, it is important to make one final comment. We have dropped additive constants when evaluating the determinant. Once these constants are reintroduced, the determinant will no longer be zero, but will no longer scale as $\det C \sim r^{-6}$, as required to cancel the dilaton shift (\ref{dilaton_shift}). Therefore, it is easy to see that additive constants, while they will contribute to a non-zero determinant, cannot help compensate the shift in the dilaton since the contribution from fermionic T-duality will appear with the wrong power of the $AdS_4$ radial direction.   

One can repeat the exercise for the other projection conditions. To retain the Killing spinors corresponding to the projector $\Gamma^{49} \eta_0 = i \eta_0$, we retain $a_1, b_1, a_5, b_5, a_6, b_6$ non-zero. Once again the constraint (\ref{C_constraint}) is trivially satisfied. The non-zero components of $C$ are 
\bea
C_{13} &=&  
\sqrt{\frac{R^3}{k}} \frac{16 \, a_1 a_5 \, i z_2 \bar{z}_3}{r (1 + |\mathbf{z}|^2)}, \quad C_{16} = \sqrt{\frac{R^3}{k}} \frac{16 \, a_1 b_6 \, i z_2 \bar{z}_1}{r (1 + |\mathbf{z}|^2)}, \nn
C_{24} &=& \sqrt{\frac{R^3}{k}} \frac{16 \, b_1 b_5 \, i z_2 \bar{z}_3}{r (1 + |\mathbf{z}|^2)}, \quad C_{25} = - \sqrt{\frac{R^3}{k}} \frac{16 \, a_6 b_1 \, i z_2 \bar{z}_1}{r (1 + |\mathbf{z}|^2)}, \nn
C_{35} &=& - \sqrt{\frac{R^3}{k}} \frac{16 \, a_5 a_6 \, i z_2}{r (1 + |\mathbf{z}|^2)}, \quad C_{46} = - \sqrt{\frac{R^3}{k}} \frac{16 \, b_5 b_6 \, i z_2}{r (1 + |\mathbf{z}|^2)}, \nn
\eea
and referring to (\ref{detC}) it is once again easy to confirm that $\det C = 0$. For the basis spinors corresponding to the projection condition $\Gamma^{58} \eta_ 0 = i \eta_0$, we identify the components, 
\bea
C_{13} &=&  
-\sqrt{\frac{R^3}{k}} \frac{16 \, a_2 a_4 \, i z_3 \bar{z}_2}{r (1 + |\mathbf{z}|^2)}, ~~ C_{15} = - \sqrt{\frac{R^3}{k}} \frac{16 \, a_2 a_6 \, i z_3 \bar{z}_1}{r (1 + |\mathbf{z}|^2)}, \nn
C_{24} &=&  -\sqrt{\frac{R^3}{k}} \frac{16 \, b_2 b_4 \, i z_3 \bar{z}_2}{r (1 + |\mathbf{z}|^2)}, ~~ C_{26} = - \sqrt{\frac{R^3}{k}} \frac{16 \, b_2 b_6 \, i z_3 \bar{z}_1}{r (1 + |\mathbf{z}|^2)}, \nn
C_{36} &=& - \sqrt{\frac{R^3}{k}} \frac{16 \, a_4 b_6 \, i z_3}{r (1 + |\mathbf{z}|^2)}, ~~ C_{45} = \sqrt{\frac{R^3}{k}} \frac{16 \, a_6 b_4 \, i z_3}{r (1 + |\mathbf{z}|^2)}
\eea
while for the spinors corresponding to $\Gamma^{67} \eta_0 = i \eta_0$, we find the following components: 
\bea
C_{14} &=& \sqrt{\frac{R^3}{k}} \frac{16 \, a_3 b_4 \, i z_1 \bar{z}_2}{r (1 + |\mathbf{z}|^2)}, ~~ C_{15} = \sqrt{\frac{R^3}{k}} \frac{16 \, a_3 a_5 \, i z_1 \bar{z}_3}{r (1 + |\mathbf{z}|^2)},  \nn
C_{23} &=&  -\sqrt{\frac{R^3}{k}} \frac{16 \, a_4 b_3 \, i z_1 \bar{z}_2}{r (1 + |\mathbf{z}|^2)}, ~~ C_{26} = \sqrt{\frac{R^3}{k}} \frac{16 \, b_3 b_5 \, i z_1 \bar{z}_3}{r (1 + |\mathbf{z}|^2)}\nn
C_{35} &=& \sqrt{\frac{R^3}{k}} \frac{16 \, a_4 a_5 \, i z_1}{r (1 + |\mathbf{z}|^2)}, ~~ C_{46} = \sqrt{\frac{R^3}{k}} \frac{16 \, b_4 b_5 \, i z_1}{r (1 + |\mathbf{z}|^2)}. 
\eea
A short calculation reveals that both determinants are zero. 

\section{Effect of TsT} 
It has been suggested that a TsT transformation \cite{Lunin:2005jy} may be employed to exorcise the singularity in the dilaton shift \cite{Bakhmatov:2010fp}. This idea has considerable merit since it allows one to both deform the geometry, while at the same time, transform the supersymmetries. Moreover, we noted earlier that once we pick the fermionic isometries, there will be bosonic directions that are independent of the preserved supersymmetry, thus making these natural candidates for TsT transformation. In this section, we will demonstrate the action of a TsT transformation on the Killing spinors. We are not aware of an existing treatment in the literature.

We consider a 10D spacetime with $U(1) \times U(1)$ isometry,
\be
\dd s^2_{10} = \dd s^2_8 + e^{2 C_1} \DD \varphi_1^2 + e^{2 C_2} \DD \varphi_2^2, 
\ee
where we have defined the covariant derivatives, $ \DD \varphi^i = \dd \varphi^i + \mathcal{A}^i$. Here $C_i$ denote scalar warp factors, and $\mathcal{A}^i$ represent gauge fields, all of which depend on the 8D spacetime. We will assume for simplicity that the Killing spinors do not depend on the isometries and we will also drop the RR sector. It is straightforward, e. g. \cite{Kelekci:2014ima}, to generalise the analysis presented here. The NS sector suffices to identify the transformation on the spinors. 

We allow for an initial dilaton $\Phi$ and NS two-form $B$
\be
B = B_2 + \mathcal{B}^1 \wedge \dd \varphi_1 + \mathcal{B}^2 \wedge \dd \varphi_2 , 
\ee
where we have defined an additional two-form, $B_2$ and 2 one-forms, $\mathcal{B}^i$, which depend on the transverse 8D spacetime. 

Performing a T-duality on $\varphi_1$, a constant shift $\varphi_2 \rightarrow \varphi_2 + \lambda \varphi_1$, and a second T-duality with respect to $\varphi_1$, we find the resulting  NS sector: 
\bea
\label{Tdual_NS}
\dd \tilde{s}^2_{10} &=& \dd s^2_8 + \frac{1}{[ 1 + \lambda^2 e^{2 C_1 + 2 C_2} ]} \biggl[ e^{2 C_1} ( \DD \varphi_1 + \lambda \mathcal{B}^2)^2 \nn &+& e^{2 C_2} (\DD \varphi_2 - \lambda \mathcal{B}^1)^2 \biggr], \nn
\tilde{B} &=& B^2 + \mathcal{B}^1 \wedge \dd \varphi_1 + \mathcal{B}^2 \wedge \dd \varphi_2 + \lambda \mathcal{B}^1 \wedge \mathcal{B}^2 \nn
&-& \frac{\lambda e^{2 C_1 + 2 C_2}}{[ 1 + \lambda^2 e^{2 C_1 + 2 C_2}]} (\DD \varphi_1 + \lambda \mathcal{B}^2) \wedge (\DD \varphi_2 - \lambda \mathcal{B}^1), \nn
\tilde{\Phi} &=& \Phi - \frac{1}{2} \ln ( 1+ \lambda^2 e^{2 C_1 + 2 C_2}). 
\eea

Inserting these expressions (\ref{Tdual_NS}) into the gravitino supersymmetry variation along the transverse 8D spacetime, 
\be
\delta \Psi_{\mu}  = \nabla_{\mu} \eta  - \frac{1}{8} H_{\mu \rho \sigma} \Gamma^{\rho \sigma} \sigma^3 \eta,  
\ee
we find 
\begin{widetext}
\bea
\delta \tilde{\Psi}_{\mu} &=& \biggl[ \nabla_{\mu}   - \frac{e^{C_1}}{ 4 \sqrt{\Delta}} (\mathcal{F}^1_{\mu \nu} + \lambda \mathcal{G}^2_{\mu \nu}) \Gamma_{\varphi_1}^{~~ \nu} -  \frac{e^{C_2}}{4 \sqrt{\Delta}} (\mathcal{F}^2_{\mu \nu} - \lambda \mathcal{G}^1_{\mu \nu}) \Gamma_{\varphi_2}^{~~ \nu} - \frac{1}{8} \mathcal{H}_{\mu \rho \sigma} \Gamma^{\rho \sigma} \sigma^3 + \frac{\lambda}{2 \Delta}  \partial_{\mu} (C_1 + C_2) e^{C_1 + C_2} \Gamma^{\varphi_1 \varphi_2} \sigma^3 \nn
&+& \frac{\lambda e^{2 C_1 + C_2}}{4 \sqrt{\Delta}} ( \mathcal{F}^1_{\mu \nu} + \lambda \mathcal{G}^2_{\mu \nu}) \Gamma^{\nu \varphi_2} \sigma^3 - \frac{\lambda e^{C_1 + 2 C_2}}{4 \sqrt{\Delta}} ( \mathcal{F}^2_{\mu \nu} - \lambda \mathcal{G}^1_{\mu \nu}) \Gamma^{\nu \varphi_1} \sigma^3 
- \frac{e^{-C_1} \sqrt{\Delta}}{4} \mathcal{G}^1_{\mu \nu}  \Gamma^{\nu \varphi_1} \sigma^3 - \frac{e^{-C_2} \sqrt{\Delta}}{4} \mathcal{G}^2_{\mu \nu}  \Gamma^{\nu \varphi_2} \sigma^3
\biggr] \tilde{\eta} \nn 
\eea
\end{widetext} 
where we have defined 
\bea
\mathcal{F}^i &=& \dd \mathcal{A}^i, ~~\mathcal{G}^i = \dd \mathcal{B}^i, ~~\mathcal{H} = \dd B_2 - \mathcal{G}^1 \wedge \mathcal{A}^1 - \mathcal{G}^2 \wedge \mathcal{A}^2, \nn \Delta &=& 1 + \lambda^2 e^{2 C_1 + 2 C_2}. 
\eea
We can now redefine 
\be
\label{rotation} 
\tilde{\eta} =  e^{-X} \eta, \quad e^{2X} = \frac{1}{\sqrt{\Delta}} ( 1+ \lambda e^{C_1 +C_2} \Gamma^{\varphi_1 \varphi_2} \sigma^3),  
\ee
to recast the supersymmetry variation in terms of the original variation: 
\bea
\delta {\Psi}_{\mu} &=& e^{X} \delta \tilde{\Psi}_{\mu}. 
\eea

The transformation on the Killing spinor under a TsT transformation is given by  (\ref{rotation}). We will now see how this transformation affects fermionic T-duality. This will allow us to show that given a geometry with a global $U(1) \times U(1)$ symmetry, a deformation based on TsT, provided it does not break supersymmetry, does not affect the determination of the matrix $C_{ij}$. As a result, we can conclude that TsT transformations, assuming they can be performed, can not remove a singularity we encounter in the dilaton shift.  

To do so, we replace $\sigma^3$ with $\Gamma_{11}$, which makes the notation consistent with (\ref{C_diff}) and (\ref{C_constraint}). We next define $ \theta = \cos^{-1} (1/\sqrt{\Delta})$, so that 
\be
\tilde{\eta} = e^{- \frac{\theta}{2} \Gamma_{\varphi_1 \varphi_2} \Gamma_{11} } \eta. 
\ee
It then follows that 
\bea
\label{eq1} 0 &=&  \bar{\eta}_i \Gamma^{\mu} \eta_j, \quad \partial^{\mu} C_{ij} = \bar{\epsilon}_i \Gamma^{\mu} \Gamma_{11} \epsilon_j, \\
\label{eq2} 0 &=&  \cos \theta \bar{\eta}_i \Gamma^{\varphi_k} \eta_j - \epsilon_{k l} \sin \theta  \bar{\eta}_i \Gamma^{\varphi_l} \Gamma_{11} \eta_j, \\
\label{eq3} \partial^{\varphi_k} C_{ij} &=& \bar{\epsilon}_i \Gamma^{\varphi_k} \Gamma_{11} \eta_j - \epsilon_{k l} \sin \theta  \bar{\eta}_i \Gamma^{\varphi_l} \eta_j, 
\eea
where $\epsilon_{12} = 1$ and $\mu \neq \varphi_i$. It is clear from (\ref{eq1}) that the transformation has not affected the determination of the matrix $C_{ij}$ in the transverse 8D spacetime. To see that it also does not affect $C_{ij}$ in the $\varphi_i$-directions, we can combine the equations (\ref{eq2}) and (\ref{eq3}) to get 
\be
\partial^{\varphi_k} C_{ij} = \sqrt{\Delta} \bar{\eta}_i \Gamma^{\varphi_k} \Gamma_{11} \eta_j. 
\ee
Taking into account the factor of $\sqrt{\Delta}$ in the deformed metric (\ref{Tdual_NS}), we come to the conclusion that the equations to be solved to determine $C_{ij}$ are invariant under TsT.

\section{T-duality and the Sigma Model}

We conclude with a few remarks concerning self-duality  of $AdS_4\times\mathbb CP^3$ in its sigma model representation. The  $AdS_4\times \mathbb{C}P^3$ background can be described by the supercoset \cite{Stefanski:2008ik, Arutyunov:2008if,Gomis:2008jt}
\be\label{eq: ads4cp3cosetsigmamodelbackground}
\frac{\text{OSp}(6|2,2)}{\text{SO}(1,3)\times\text{U}(3)}  = AdS_4\times \mathbb{C}P^3 \,+\,\text{24 ferm}.
\ee
If $n\in\mathbb N^*$, a  basis of $\mathfrak{osp}(2n|2,2)$  convenient for T-duality is $
\mathfrak{osp}(2n|2,2)=\text{span}\acomm{P_{\alpha\beta},K_{\alpha\beta},D,J_{\alpha\beta},L^\pm_{AB},R_{AB}\,|\,Q_{A\alpha},\bar Q_{A\alpha},S_{A\alpha},\bar S_{A\alpha}}$, $\alpha,\beta=1,2;\;A,B=1,\dots,n$. 
The generators  above satisfy the graded commutation relations reported in the appendix and are such that 
\bea
 P_{\alpha\beta}=P_{(\alpha\beta)} &\to& \text{3 components},\nn
 K_{\alpha\beta} =K_{(\alpha\beta)} &\to& \text{3 components},\nn
 J_{\alpha\alpha}=0 &\to& \text{3 components}, \nn D &\to& \text{1 component},\nn R_{AB} &\to& n^2\text{ components},\nn
L^+_{AB}=L^+_{[AB]} &\to& n(n-1)/2\text{ components},\nn L^-_{AB}=L^-_{[AB]} &\to& n(n-1)/2 \text{ components},\nn
Q_{A\alpha},\; S_{A\alpha}, \; \bar Q_{A\alpha},\; \bar S_{A\alpha} &\to& 2n \text{ components each}.
\eea
As a consequence, $P_{\a\b},K_{\a\b},D,J_{\alpha\beta}$ respectively encode translations, special conformal transformations, dilatations and Lorentz rotations on the three-dimensional conformal  boundary of $AdS_4$. Furthermore, $L^\pm_{AB},R_{AB}$ span the SO($2n$) R-symmetry of  $\mathfrak{osp}(2n|4)$, while $Q,S,\bar Q,\bar S$ are the supercharges related to the $\mathcal N=2n$ supersymmetry of the boundary  theory. Specifically, the case $n=3$  corresponds to ABJM theory.

Writing the  coset action requires an order 4 automorphism  of $\mathfrak g=\mathfrak{osp}(2n|2,2)$, $\Omega$,  providing the projectors
\be
\mathscr P_{(k)}=\f14\p{\mathbbm 1+i^{3k}\,\Omega+i^{2k}\,\Omega^2+i^k\,\Omega^3}, ~ k=0,\dots,3.
\ee
Such projectors split the superalgebra $\mathfrak g$ into the direct sum of $\Omega$ eigenspaces:
\bea
&&\mathfrak g=\bigoplus_{k=0}^3\mathfrak g_{(k)},\quad \mathfrak g_{(k)}:=\acomm{\mathfrak J\in\mathfrak g:\Omega\p{\mathfrak J}=i^k\,\mathfrak J},\nn 
&&\gcomm{\mathfrak g_{(k)},\mathfrak g_{(l)}} \subset \mathfrak g_{(k+l)\text{ mod } 4},
\eea
where $\gcomm{\cdot,\cdot}$ is the graded Lie bracket of $\mathfrak g$. The eigenspace $\mathfrak g_{(0)}$ is a closed subalgebra of $\mathfrak g$ and the desired coset is the quotient of the exponential maps of $\mathfrak g$ and $\mathfrak g_{(0)}$, namely $\text{Exp}\p{\mathfrak g}/\text{Exp}\p{\mathfrak g_{(0)}}$. The lagrangian of the model is obtained by picking up a coset representative $g:\Sigma\to \text{Exp}\p{\mathfrak g}$, where $\Sigma$ is the string worldsheet, and decomposing the related Cartan-Maurer one-form $j=g^{-1} \dd g$ according to the $\mathbb Z_4$ grading:
\be\label{eq: maurercartancurrentdecomposition}
j=g^{-1} \dd g=j_{(0)}+j_{(1)}+j_{(2)}+j_{(3)},\qquad j_{(k)}\in \mathfrak g_{(k)}.
\ee
Finally, the action of the $\text{Exp}\p{\mathfrak g}/\text{Exp}\p{\mathfrak g_{(0)}}$ sigma model reads
\be\label{eq: greenschwarzcosetaction}
S=-(T/2)\int_\Sigma j_{(2)}\wedge*j_{(2)}+\kappa\,j_{(1)}\wedge j_{(3)},
\ee
with $T$ being the string tension and $\kappa$ the kappa-symmetry parameter \footnote{Integrability sets  $\kappa=\pm1$.}.

In general, the automorphism  $\Omega$ acts on the supercharges as \footnote{The matrices $\omega_{AB},\sigma_{\alpha\beta}$ need to be skew-symmetric in order to mode out the correct subgroup in the coset, as we shall see.}
\bea
\Omega\p{Q_{A\alpha}}=iS_{B\beta}\,\omega_{BA}\,\sigma_{\beta\alpha}, ~ \Omega\p{S_{A\alpha}}=iQ_{B\beta}\,\omega_{BA}\,\sigma_{\beta\alpha},
\eea
and similarly for $\bar Q,\bar S$. The matrix  $\omega_{AB}=\omega_{[AB]}$ and $\sigma_{\a\b}=\sigma_{[\a\b]}$ fulfill 
\be
\omega_{AC}\,\omega_{BC}=\delta_{AB},\qquad   \sigma_{\alpha\gamma}\,\sigma_{\beta\gamma}=\delta_{\alpha\beta},
\ee
and the corresponding projections of the supercharges read 
\ale{
\mathfrak Q^{1,3}_{A\alpha}:=\mathscr P_{1,3}\,Q_{A\alpha}&= \f14\p{Q_{A\alpha}\pm S_{B\beta}\,\omega_{BA}\sigma_{\beta\alpha}}, \nn \bar {\mathfrak Q}^{1,3}_{A\alpha}:=\mathscr P_{1,3}\,\bar Q_{A\alpha}&=\f14\p{ \bar Q_{A\alpha}\pm\bar S_{B\beta}\,\omega_{BA}\sigma_{\beta\alpha}}.
}
The complete $ \mathbb Z_4$ grading of $\mathfrak g$ is:
\begin{widetext}
\ale{
\mathfrak g_{0}&=\bk{\delta_{AB}\,P_{\alpha\beta}+{\sigma_{\alpha\gamma}\,K_{\gamma\delta}\,\sigma_{\delta\beta}},\; J_{\alpha\beta},\; L^+_{AD}\,\omega_{DB}+L^-_{BD}\,\omega_{DA},\;\omega_{C(B}R_{A)C}}\nn
\mathfrak g_{1}&=\bk{Q_{A\alpha}+S_{B\beta}\,\omega_{BA}\sigma_{\beta\alpha},\; \bar Q_{A\alpha}+\bar S_{B\beta}\,\omega_{BA}\sigma_{\beta\alpha}}\nn
\mathfrak g_{2}&=\bk{\delta_{AB}\,P_{\alpha\beta}-{\sigma_{\alpha\gamma}\,K_{\gamma\delta}\,\sigma_{\delta\beta}},\; D,\; L^+_{AD}\,\omega_{DB}-L^-_{BD}\,\omega_{DA},\;\omega_{C[B}R_{A]C}}\nn
\mathfrak g_{3}&=\bk{Q_{A\alpha}-S_{B\beta}\,\omega_{BA}\sigma_{\beta\alpha},\; \bar Q_{A\alpha}-\bar S_{B\beta}\,\omega_{BA}\sigma_{\beta\alpha}}.
}
\end{widetext}
 
If $n=3$, $\mathfrak g= \text{OSp}(6|2,2)$, $\mathfrak g_0= \mathfrak{so}(1,3)\oplus \mathfrak u(3) $ and the coset is
\be
\f{\text{OSp}(6|2,2)}{\text{SO}(1,3)\times \text{U}(3) }\ =\ \f{\text{Sp}(2,2)}{\text{SO}(1,3)}\times \f{\text{SO}(6)}{ \text{U}(3) }\,+\,\text{24 ferm.},
\ee
which is exactly (\ref{eq: ads4cp3cosetsigmamodelbackground}). Therefore, one chooses the coset representative
\bea
g &=& e^{X_{\beta\alpha} \,P_{\alpha\beta}\, + \, \lambda_{BA}\, L^+_{AB}\,+\,\theta_{A\alpha} \,Q_{A\alpha} }\times \\ &\times& e^{-\,\bar \theta_{A\alpha}\, \bar Q_{A\alpha}\,-\,\bar\xi_{A\alpha}\, \bar S_{A\alpha}}\,e^{-\,D\log Y-\,\rho_{BA}\,R_{AB}}\,e^{-\xi_{A\alpha} \,S_{A\alpha} }, \nonumber
\eea 
finds the current components (\ref{eq: maurercartancurrentdecomposition}) and the action (\ref{eq: greenschwarzcosetaction}). T-duality for backgrounds with  isometry supergroup of OSp type \cite{Abbott:2015mla} requires to apply Buscher rules on $P_{\a\b},L^+_{AB},Q_{A\alpha}$ , which are 3 bosonic directions along the conformal boundary of $AdS_4$, 3 bosonic directions along $\mathbb CP^3$ and 6 fermionic directions  respectively. T-duality maps these  as follows:
\be
\bk{P_{\a\b},L^+_{AB},Q_{A\a}}\quad \longrightarrow\quad  \bk{K_{\a\b},L^-_{AB},S_{A\a}}.
\ee
In particular, T-duality along $\theta$ and $\lambda$ inverts the metrics of the corresponding kinetic terms. Unfortunately, these metrics contain the matrix $\omega_{AB}$, which for OSp(6$|$2,2) is a $3\times3$ skew-symmetric matrix and, as such, is not invertible. This is a direct consequence of the fact that OSp(6$|$2,2) does not admit a non-singular outer automorphism of order four \cite{  Arutyunov:2008if,Gomis:2008jt}.

On the other hand,  $AdS_4\times \mathbb CP^3$ can be embedded into a bigger system. Indeed, if $n=4$, $\mathfrak g= \text{OSp}(8|2,2)$, $\mathfrak g_0= \mathfrak{so}(1,3)\oplus \mathfrak u(4) $, and the coset becomes
\bea\label{eq: embeddingbackground}
&& \f{\text{OSp}(8|2,2)}{\text{SO}(1,3)\times \text{U}(4) }\ =\ \f{\text{Sp}(2,2)}{\text{SO}(1,3)}\times \f{\text{SO}(8)}{ \text{SO}(2)\times \text{SO}(6) } \nn &&+\,\text{32 ferm.}=\ AdS_4 \times G_2\p{\mathbb {R}^8}+\,\text{32 ferm}. 
\eea
The dimension of the Gra\ss mannian $G_2(\mathbb {R}^8)$ is 12 and the bosonic dimension of the supercoset (\ref{eq: embeddingbackground}) is  16. This is not a string background \fn{This model, which is defined on a projective superspace, should be related to those studied in \cite{Candu:2009ep}.}, but it contains $AdS_4\times \mathbb CP^3$ and can be used to study the action of T-duality upon the latter. Indeed, Buscher rules are not singular for $AdS_4 \times G_2\p{\mathbb {R}^8}$ because the skew-symmetric matrix appearing in the fermionic  kinetic terms, $\omega_{AB}$, is now $4\times4$ and invertible \fn{The role of  $\omega_{AB}$ can be played by a Sp(4) metric, for instance.}. The  coset given in (\ref{eq: embeddingbackground}) can therefore be used to map a $AdS_4\times \mathbb CP^3$ submanifold of (\ref{eq: embeddingbackground})  into a dual $AdS_4\times \mathbb CP^3$ submanifold.   Notice that the Berezinian of the transformation is non-trivial,
\bea\label{eq: nontrivialberezinian}
\p{2\times\#_P-\#_Q}\log r =\p{2\times 3-8}\log r \neq0.
\eea
The super-Jacobian of the transformation is not 1 and the measure of the corresponding path integral would not be left  unchanged by  the Buscher procedure just described.  As a consequence,  the mapping between different $AdS_4 \times {\mathbb {C}}P^3$ slices of of  $AdS_4 \times G_2\p{\mathbb {R}^8}$  can  only be understood as a \emph{classical} symmetry of  the model, not as a quantum one. 

 In summary: by writing the $AdS_4 \times \mathbb  CP^3$ sigma model action in the patch proper to perform T-duality (borrowed from  \cite{Abbott:2015mla}), we found  unavoidable singularities arising from the kinetic terms of the fermions and of the $\mathbb  CP^3$ coordinates  that are affected by Buscher procedure. The reason for these singularities is group theoretical, as it descends from the fact that OSp(6$|$2,2) does not admit an invertible outer automorphism of order 4.  Moreover, we showed that  $AdS_4 \times \mathbb  CP^3$ can be embedded into $AdS_4 \times G_2\p{\mathbb {R}^8}$, which is classically self-dual under a combination of T-dualities. In particular, T-duality maps to each other different $AdS_4 \times \mathbb  CP^3$ slices of  $AdS_4 \times G_2\p{\mathbb {R}^8}$. As already mentioned, this self-duality has a clear interpretation only at the classical level; thus, it cannot justify the dual superconformal symmetry of ABJM theory.

\section{Conclusions} 
In this letter we have studied commuting bosonic and fermionic isometries for the geometry $AdS_4 \times \mathbb{C} P^3$ in a systematic way to address if it is self-dual with respect to a combination of T-dualities. Employing both supergravity and sigma model analysis, we demonstrated that irrespective of the chosen isometries, one encounters a singularity in the dilaton shift. While TsT transformations provide a natural way to deform the geometry and still preserve supersymmetry, we show that it commutes with fermionic T-duality, and so will not affect our conclusions. 

We remark that fermionic T-duality has been derived from a supergravity ansatz as a special case of a more general transformation involving Killing spinor bilinears \cite{Godazgar:2010ph}, where $C_{ij}$ may include anti-symmetric components. However, (\ref{C_constraint}) is also a constraint for this generalisation, and as we have worked with the explicit Killing spinors, it is not clear how $C_{ij}$ may possess an anti-symmetric part in the current setting. Furthermore, one may imagine that the singularity could be resolved by lifting the problem to 11D supergravity, but it is worth recalling that the perturbative evidence for self-duality holds in the IIA regime. 

While our results preclude self-duality based on fermionic T-duality, the wealth of perturbative results, some of which are connected to known integrable structures, i. e. Yangian, suggests that some self-duality transformation should be at work. In this light, it is important to understand the connection between integrability and self-duality. $AdS_4 \times \mathbb{C} P^3$ aside, it is clear that the Berkovits-Maldacena transformation, since it relies on preserved supersymmetry, can not be responsible for self-duality for quotients and TsT deformations of $AdS_5 \times S^5$, despite the presence of integrability (see for example \cite{Zoubos:2010kh}). As supersymmetry is decreased, our useful rule of thumb means we can not find the requisite number of fermionic isometries required to undo the dilaton shift from the anti-de Sitter T-dualities. 

In fact, the Lunin-Maldacena solutions \cite{Lunin:2005jy} are self-dual, as all one has to do is undo the TsT transformation, apply self-duality and re-apply TsT. Through this chain of dualities, it is clear that TsT-deformed $AdS_5 \times S^5$ can be self-dual, but there should be a generalistion of fermionic T-duality that holds directly when supersymmetry is broken.  We plan to pursue this in future work in the hope that it sheds some light on the expected self-duality of $AdS_4 \times \mathbb{C} P^3$.

\section*{Acknowledgements}
We are grateful I. Bakhmatov, S. Lee, M. Wolf for comments on earlier drafts and M. Abbott, C. Meneghelli, S. Penati, A. Prinsloo, W. Siegel, D. Sorokin, A. Torrielli \& L. Wulff for discussion. E. \'O C is grateful to the Simons Center for Geometry \& Physics for hospitality and is supported by the fellowship PIOF-2012- 328625 T-DUALITIES. A. P. was supported in part by the EPSRC under the grant EP/K503186/1. 

\section*{Data Management} 
No additional research data beyond the data presented and cited in this work are needed to validate the research findings in this work. 

\appendix
\section{Gamma matrices} 
In this work we make use of the following real gamma matrices, 
\bea
\Gamma^0 = i \sigma^2 \otimes 1_{16}, \quad \Gamma^{i} = \sigma^1 \otimes \gamma^i, 
\eea
where 
\bea
\gamma_1 &=& \sigma^2 \otimes \sigma^2 \otimes \sigma^2 \otimes \sigma^2, \quad 
\gamma_2 = \sigma^2 \otimes 1_2 \otimes \sigma^1 \otimes \sigma^2, \nn
\gamma_3 &=& \sigma^2 \otimes 1_2 \otimes \sigma^3 \otimes \sigma^2, \quad
\gamma_4 = \sigma^2 \otimes \sigma^1 \otimes \sigma^2 \otimes 1_2, \nn
\gamma_5 &=& \sigma^2 \otimes \sigma^3 \otimes \sigma^2 \otimes 1_2, \quad
\gamma_6 = \sigma^2 \otimes \sigma^2 \otimes 1_2 \otimes \sigma^1, \nn
\gamma_7 &=& \sigma^2 \otimes \sigma^2 \otimes 1_2 \otimes \sigma^3, \quad
\gamma_8 = \sigma^1 \otimes 1_{2} \otimes 1_2 \otimes 1_2 , \nn
\gamma_9 &=& \sigma^3 \otimes 1_{2} \otimes 1_2 \otimes 1_2. 
\eea
Observe with this representation that $\Gamma_{11} \equiv \Gamma^{0123456789} = \sigma^3 \otimes 1_{16}$, $\Gamma^0$ is anti-symmetric, while $\Gamma^i$ are symmetric. 

Using the above gamma matrices, we can construct explicit basis spinors (\ref{basis_spinor}), 
\begin{widetext}
\bea
\xi_{+--} &=& \left( \begin{array}{c} a_1 \\ b_1 \end{array} \right) \otimes  \left( \begin{array}{c} 1 \\ 0 \\0 \\ 1 \end{array} \right) \otimes \left( \begin{array}{c} 1 \\ i \\ i \\ -1 \end{array}\right)  + 
 \left( \begin{array}{c} b_1 \\ -a_1 \end{array} \right) \otimes \left( \begin{array}{c} 0 \\ 1 \\ -1 \\ 0 \end{array} \right) \otimes \left( \begin{array}{c} 1 \\i \\ -i \\ 1  \end{array}\right), \nn
\xi_{-+-} &=& \left( \begin{array}{c} a_2 \\ b_2 \end{array} \right) \otimes  \left( \begin{array}{c} 1 \\ 0 \\0 \\ 1 \end{array} \right) \otimes \left( \begin{array}{c} 1 \\ i \\ -i \\ 1 \end{array}\right)  + 
 \left( \begin{array}{c} b_2 \\ -a_2 \end{array} \right) \otimes \left( \begin{array}{c} 0 \\ 1 \\ -1 \\ 0 \end{array} \right) \otimes \left( \begin{array}{c} 1 \\i \\ i \\ -1  \end{array}\right), \nn
\xi_{--+} &=& \left( \begin{array}{c} a_3 \\ b_3 \end{array} \right) \otimes  \left( \begin{array}{c} 1 \\ 0 \\0 \\ -1 \end{array} \right) \otimes \left( \begin{array}{c} 1 \\ -i \\ i \\ 1 \end{array}\right)  + 
 \left( \begin{array}{c} b_3 \\ -a_3 \end{array} \right) \otimes \left( \begin{array}{c} 0 \\ 1 \\ 1 \\ 0 \end{array} \right) \otimes \left( \begin{array}{c} 1 \\-i \\ -i \\ -1  \end{array}\right), \nn
\xi_{-++} &=& \left( \begin{array}{c} a_4 \\ b_4 \end{array} \right) \otimes  \left( \begin{array}{c} 1 \\ 0 \\0 \\ 1 \end{array} \right) \otimes \left( \begin{array}{c} 1 \\ -i \\ -i \\ -1 \end{array}\right)  +  \left( \begin{array}{c} b_4 \\ -a_4 \end{array} \right) \otimes \left( \begin{array}{c} 0 \\ 1 \\ -1 \\ 0 \end{array} \right) \otimes \left( \begin{array}{c} 1 \\ -i \\ i \\ 1  \end{array}\right), \nn
\xi_{+-+} & = & \left( \begin{array}{c} a_5 \\ b_5 \end{array} \right) \otimes  \left( \begin{array}{c} 1 \\ 0 \\0 \\ 1 \end{array} \right) \otimes \left( \begin{array}{c} 1 \\ -i \\ i \\ 1 \end{array}\right)  +  \left( \begin{array}{c} b_5 \\ -a_5 \end{array} \right) \otimes \left( \begin{array}{c} 0 \\ 1 \\ -1 \\ 0 \end{array} \right) \otimes \left( \begin{array}{c} 1 \\ -i \\ -i \\ -1  \end{array}\right), \nn
\xi_{++-} &=& \left( \begin{array}{c} a_6 \\ b_6 \end{array} \right) \otimes  \left( \begin{array}{c} 1 \\ 0 \\0 \\ -1 \end{array} \right) \otimes \left( \begin{array}{c} 1 \\ i \\ -i \\ 1 \end{array}\right)  + 
 \left( \begin{array}{c} b_6 \\ -a_6 \end{array} \right) \otimes \left( \begin{array}{c} 0 \\ 1 \\ 1 \\ 0 \end{array} \right) \otimes \left( \begin{array}{c} 1 \\i \\ i \\ -1  \end{array}\right), \eea
\end{widetext}
where $a_i, b_i$ are complex constants. Note, the first three and last three spinors are modulo constants, complex conjugates, as expected. 

\newpage 

\section{Kosmann Derivatives} 
In this section, we record the Kosmann derivatives for various vectors. For the vectors lengthy, but straightforward calculations reveal: 
\begin{widetext}
\bea
\mathcal{L}_{K_1} \eta &=& \frac{1}{4} e^{ \frac{i}{2} ( \phi -\chi)} e^{\frac{\mu}{2} (\Gamma^9 i \sigma^2 + \Gamma^{43})} e^{- \frac{\alpha}{2} (\Gamma^{54}+\Gamma^{89})} \biggl( e^{ \frac{i}{2} \psi} \cos \frac{\theta}{2} ( i \Gamma^7 - \Gamma^6) \nn &+& e^{-\frac{i}{2} \psi}  \sin \frac{\theta}{2} (i \Gamma^8-\Gamma^5) \biggr) [\sigma^2 - \Gamma^3] e^{\frac{\alpha}{2} (\Gamma^{54} + \Gamma^{89})}  e^{-\frac{\mu}{2} (\Gamma^9 i \sigma^2 + \Gamma^{43})}  \eta, 
\eea
\bea
 \mathcal{L}_{K_2} \eta &=& \frac{1}{4} e^{-\frac{i}{2} \chi} e^{\frac{\mu}{2} (\Gamma^9 i \sigma^2 + \Gamma^{43})} e^{- \frac{\alpha}{2} (\Gamma^{54}+\Gamma^{89})}(i \Gamma^9 - \Gamma^4) [\sigma^2 - \Gamma^3 ] e^{\frac{\alpha}{2} (\Gamma^{54} + \Gamma^{89})}  e^{-\frac{\mu}{2} (\Gamma^9 i \sigma^2 + \Gamma^{43})}  \eta, 
 \eea
 \bea
\mathcal{L}_{K_3} \eta &=& \frac{1}{4} e^{- \frac{i}{2} ( \phi +\chi)} e^{\frac{\mu}{2} (\Gamma^9 i \sigma^2 + \Gamma^{43})} e^{- \frac{\alpha}{2} (\Gamma^{54}+\Gamma^{89})} \biggl( e^{- \frac{i}{2} \psi} \cos \frac{\theta}{2} ( i \Gamma^8 - \Gamma^5) \nn &-& e^{\frac{i}{2} \psi}  \sin \frac{\theta}{2} (i \Gamma^7-\Gamma^6) \biggr) [\sigma^2 - \Gamma^3] e^{\frac{\alpha}{2} (\Gamma^{54} + \Gamma^{89})}  e^{-\frac{\mu}{2} (\Gamma^9 i \sigma^2 + \Gamma^{43})}  \eta, 
\eea
\bea
\mathcal{L}_{K_{11}} &=& -\frac{i}{2}  e^{\frac{\mu}{2} (\Gamma^9 i \sigma^2 + \Gamma^{43})} e^{- \frac{\alpha}{2} (\Gamma^{54}+\Gamma^{89})}  e^{\frac{\chi}{4} (\Gamma^{67}+\Gamma^{58}+ \Gamma^{49})}e^{\frac{\psi}{4} (\Gamma^{58}-\Gamma^{67})} e^{- \frac{\theta}{4} (\Gamma^{65}+\Gamma^{78})} \Gamma^{67} \times \nn
&&e^{\frac{\theta}{4} (\Gamma^{65}+\Gamma^{78})}  e^{-\frac{\psi}{4} (\Gamma^{58}-\Gamma^{67})} e^{-\frac{\chi}{4} (\Gamma^{67}+\Gamma^{58}+ \Gamma^{49})} e^{\frac{\alpha}{2} (\Gamma^{54}+\Gamma^{89})} e^{-\frac{\mu}{2} (\Gamma^9 i \sigma^2 + \Gamma^{43})} \eta, 
\eea
\bea
\mathcal{L}_{K_{22}} &=& -\frac{i}{2}  e^{\frac{\mu}{2} (\Gamma^9 i \sigma^2 + \Gamma^{43})} e^{- \frac{\alpha}{2} (\Gamma^{54}+\Gamma^{89})}  e^{\frac{\chi}{4} (\Gamma^{67}+\Gamma^{58}+ \Gamma^{49})}e^{\frac{\psi}{4} (\Gamma^{58}-\Gamma^{67})} e^{- \frac{\theta}{4} (\Gamma^{65}+\Gamma^{78})} \Gamma^{49} \times \nn
&&e^{\frac{\theta}{4} (\Gamma^{65}+\Gamma^{78})}  e^{-\frac{\psi}{4} (\Gamma^{58}-\Gamma^{67})} e^{-\frac{\chi}{4} (\Gamma^{67}+\Gamma^{58}+ \Gamma^{49})} e^{\frac{\alpha}{2} (\Gamma^{54}+\Gamma^{89})} e^{-\frac{\mu}{2} (\Gamma^9 i \sigma^2 + \Gamma^{43})} \eta, 
\eea
\bea
\mathcal{L}_{K_{33}} &=& -\frac{i}{2}  e^{\frac{\mu}{2} (\Gamma^9 i \sigma^2 + \Gamma^{43})} e^{- \frac{\alpha}{2} (\Gamma^{54}+\Gamma^{89})}  e^{\frac{\chi}{4} (\Gamma^{67}+\Gamma^{58}+ \Gamma^{49})}e^{\frac{\psi}{4} (\Gamma^{58}-\Gamma^{67})} e^{- \frac{\theta}{4} (\Gamma^{65}+\Gamma^{78})} \Gamma^{58} \times \nn
&&e^{\frac{\theta}{4} (\Gamma^{65}+\Gamma^{78})}  e^{-\frac{\psi}{4} (\Gamma^{58}-\Gamma^{67})} e^{-\frac{\chi}{4} (\Gamma^{67}+\Gamma^{58}+ \Gamma^{49})} e^{\frac{\alpha}{2} (\Gamma^{54}+\Gamma^{89})} e^{-\frac{\mu}{2} (\Gamma^9 i \sigma^2 + \Gamma^{43})} \eta, 
\eea
\bea
\mathcal{L}_{K_{31}} \eta &=& - \frac{1}{4}  e^{i \phi} e^{\frac{\mu}{2} (\Gamma^9 i \sigma^2 + \Gamma^{43})} e^{- \frac{\alpha}{2} (\Gamma^{54}+\Gamma^{89})} e^{\frac{\psi}{4} (\Gamma^{58}-\Gamma^{67})} e^{- \frac{\theta}{4} (\Gamma^{65}+\Gamma^{78})} [ \Gamma^{78} + \Gamma^{65} + i (\Gamma^{68} + \Gamma^{57}) ] \times \nn 
&& e^{\frac{\theta}{4} (\Gamma^{65}+\Gamma^{78})}  e^{-\frac{\psi}{4} (\Gamma^{58}-\Gamma^{67})} e^{\frac{\alpha}{2} (\Gamma^{54}+\Gamma^{89})} e^{-\frac{\mu}{2} (\Gamma^9 i \sigma^2 + \Gamma^{43})} \eta, \\
\mathcal{L}_{K_{12}} \eta &=& \frac{1}{4} e^{\frac{i}{2} (\psi - \phi)} e^{\frac{\mu}{2} (\Gamma^9 i \sigma^2 + \Gamma^{43})} e^{- \frac{\alpha}{2} (\Gamma^{54}+\Gamma^{89})} \biggl[ \cos \frac{\theta}{2} e^{-i \psi} [ (\Gamma^{79}-\Gamma^{46}) - i (\Gamma^{47} + \Gamma^{69})]\nn &-& \sin \frac{\theta}{2} [ \Gamma^{45} - \Gamma^{89} + i(\Gamma^{59}+ \Gamma^{48})] \biggr] e^{\frac{\alpha}{2} (\Gamma^{54}+\Gamma^{89})} e^{-\frac{\mu}{2} (\Gamma^9 i \sigma^2 + \Gamma^{43})} \eta, 
\eea
\bea
\mathcal{L}_{K_{23}} \eta &=& \frac{1}{4} e^{-\frac{i}{2} (\psi + \phi)} e^{\frac{\mu}{2} (\Gamma^9 i \sigma^2 + \Gamma^{43})} e^{- \frac{\alpha}{2} (\Gamma^{54}+\Gamma^{89})} \biggl[ \cos \frac{\theta}{2}  [ \Gamma^{45}-\Gamma^{89} - i (\Gamma^{59} + \Gamma^{48})]\nn &+& \sin \frac{\theta}{2} e^{i \psi} \left[ \Gamma^{79}- \Gamma^{46} + i (\Gamma^{47}+\Gamma^{69} \right] e^{\frac{\alpha}{2} (\Gamma^{54}+\Gamma^{89})} e^{-\frac{\mu}{2} (\Gamma^9 i \sigma^2 + \Gamma^{43})} \eta. 
\eea
\end{widetext}

In terms of angular coordinates, we can re-express the vectors $K_{aa}$ as 
\bea
K_{11} &=& - i (\partial_{\psi} - \partial_{\phi}), \quad K_{22} = 2 i (\partial_{\psi} - \partial_{\chi}), \nn K_{33} &=& - i (\partial_{\psi}+ \partial_{\phi}). 
\eea


\section{$\mathfrak{osp}(2n|4)$ Superconformal Algebra}

The bosonic commutation relations for $\mathfrak{osp}(2n|4)$ are \footnote{The superalgebra  $\mathfrak{osp}(2n|4)$ can be understood as the Euclidean version of $\mathfrak{osp}(2n|2,2)$. Notice that the  results of this paper are independent of the spacetime signature or of the reality conditions on the supercharges.}:
\begin{widetext}
\bea
\comm{D,P_{\alpha\beta}}&=&P_{\alpha\beta},\qquad \comm{D,K_{\alpha\beta}}=-K_{\alpha\beta},\qquad \comm{J_{\alpha\beta},J_{\gamma\delta}}=\delta_{\beta\gamma}\,J_{\alpha\delta}-\delta_{\alpha\delta}\,J_{\gamma\beta}, \nn
\comm{P_{\alpha\beta},K_{\gamma\delta}}&=&-4\,\delta_{\alpha(\gamma}\,\delta_{\delta)\beta}\,D-2\,\delta_{\gamma(\alpha}\,J_{\beta)\delta}-2\,\delta_{\delta(\alpha}\,J_{\beta)\gamma},\nn
\comm{J_{\alpha\beta},P_{\gamma\delta}}&=&2\,P_{\alpha(\gamma}\,\delta_{\delta)\beta}-\delta_{\alpha\beta}\,P_{\gamma\delta},\quad \comm{J_{\alpha\beta},K_{\gamma\delta}}=-2\,K_{\alpha(\gamma}\,\delta_{\delta)\beta}+\delta_{\alpha\beta}\,K_{\gamma\delta},\nn
\comm{R_{AB},L^+_{CD}}&=&2\,L^+_{A[D}\delta_{C]B},\qquad \comm{R_{AB},L^-_{CD}}=2\,L^-_{B[C}\delta_{D]A},\nn 
\comm{L^+_{AB},L^-_{CD}}&=&2\,R_{A[D}\,\delta_{C]B}-2\,R_{B[D}\,\delta_{C]A},\quad \comm{R_{AB},R_{CD}}=\delta_{BC}R_{AD}-\delta_{AD}R_{CB}.
\eea
\end{widetext}
The fermion-fermion anticommutators read:
\begin{widetext}
\ale{ 
\acomm{Q_{A\alpha},\bar Q_{B\beta}}&=\delta_{AB}\,P_{\alpha\beta}, \qquad \acomm{S_{A\alpha},\bar S_{B\beta}}=\delta_{AB}\,K_{\alpha\beta}\nn
\acomm{Q_{A\alpha},S_{B\beta}}&=\delta_{AB}\,\delta_{\alpha\beta}\,D+\delta_{AB}\,J_{\alpha\beta}-\delta_{\alpha\beta}\,R_{AB},\qquad \acomm{Q_{A\alpha},\bar S_{B\beta}}=-\delta_{\alpha\beta}\,L^+_{AB}\nn
\acomm{\bar Q_{A\alpha},\bar S_{B\beta}}&=\delta_{AB}\,\delta_{\alpha\beta}\,D+\delta_{AB}\,J_{\alpha\beta}+\delta_{\alpha\beta}\,R_{BA},\qquad \acomm{\bar Q_{A\alpha}, S_{B\beta}}=-\delta_{\alpha\beta}\,L^-_{AB}.
}
\end{widetext}
The boson-fermion commutators are:
\begin{widetext}
\ale{ 
\comm{D,Q_{A\alpha}}&=\f12\,Q_{A\alpha}, \quad \comm{D,S_{A\alpha}}=-\f12\,S_{A\alpha},\quad \comm{D,\bar Q_{A\alpha}}=\f12\,\bar Q_{A\alpha}, \quad \comm{D,\bar S_{A\alpha}}=-\f12\,\bar S_{A\alpha}\nn
\comm{J_{\alpha\beta},Q_{A\gamma}}&=\delta_{\beta\gamma}\,Q_{A\alpha}-\f12\,\delta_{\alpha\beta}\,Q_{A\gamma},\quad \comm{J_{\alpha\beta},S_{A\gamma}}=-\delta_{\alpha\gamma}\,S_{A\beta}+\f12\,\delta_{\alpha\beta}\,S_{A\gamma}\nn
\comm{P_{\alpha\beta},S_{A\gamma}}&=2\,\bar Q_{A(\alpha}\,\delta_{\beta)\gamma},\quad \comm{K_{\alpha\beta},Q_{A\gamma}}=2\,\bar S_{A(\alpha}\,\delta_{\beta)\gamma}\nn
\comm{R_{AB},Q_{C\alpha}}&=\delta_{BC} \, Q_{A\alpha},\quad \comm{R_{AB},S_{C\alpha}}=-\delta_{AC} \, S_{B\alpha},\nn
\comm{L^+_{AB},S_{C\alpha}}&=2\,\delta_{C[B}\,\bar S_{A]\alpha},\quad \comm{L^-_{AB},Q_{C\alpha}}=2\,\delta_{C[B}\,\bar Q_{A]\alpha}
}
\end{widetext}
and the Killing forms are 
\begin{widetext}
\ale{
\str\p{P_{\alpha\beta}\,K_{\gamma\delta}}&=-4\,\delta_{\alpha(\gamma}\,\delta_{\delta)\beta},\quad \str\p{DD}=1,\quad \str\p{J_{\alpha\beta}\,J_{\gamma\delta}}=2\,\delta_{\alpha\delta}\,\delta_{\beta\gamma}-\delta_{\alpha\beta}\,\delta_{\gamma\delta}\nn
\str\p{L^+_{AB}\,L^-_{CD}}&=4\,\delta_{A[D}\,\delta_{C]B},\quad \str\p{R_{AB}\,R_{CD}}=2\,\delta_{AD}\,\delta_{BC},\nn
\str\p{Q_{A\alpha}\,S_{B\beta}}&=\delta_{AB}\,\delta_{\alpha\beta},\quad \str\p{\bar Q_{A\alpha}\,\bar S_{B\beta}}=\delta_{AB}\,\delta_{\alpha\beta}.
}
\end{widetext}
Here, $\alpha,\beta,\gamma,\delta=1,2$ while $A,B,C,D=1,\dots,n$.

\end{document}